\documentclass[10pt, onecolumn, conference]{IEEEtran}

\usepackage[usenames,dvipsnames]{color}
\usepackage{graphicx,url}
\usepackage{caption}
\usepackage[utf8]{inputenc}
\usepackage{tabularx}
\usepackage{multirow}
\usepackage{subcaption}
\usepackage[portuguese, ruled, linesnumbered]{algorithm2e}
\usepackage{multicol}
\usepackage{amssymb}
\usepackage{pifont}
\usepackage{setspace}
\usepackage{geometry}
\usepackage[english]{babel}

\usepackage{booktabs}       

\usepackage{enumitem}
\usepackage{etoolbox}
\newbool{showSilvioChanges}
\setbool{showSilvioChanges}{true}
\definecolor{myblue}{RGB}{0, 83, 216}
\newcommand{\seq}[1]{\ifbool{showSilvioChanges}{{\color{myblue}#1}}{#1}}
\usepackage{array}
\newcolumntype{L}[1]{>{\raggedright\arraybackslash}p{#1}}
\newcolumntype{C}[1]{>{\centering\arraybackslash}p{#1}}
\newcolumntype{R}[1]{>{\raggedleft\arraybackslash}p{#1}}

\newbool{showVagnerChanges}
\setbool{showVagnerChanges}{true}
\definecolor{myColorV}{RGB}{178, 113, 51}
\newcommand{\veq}[1]{\ifbool{showVagnerChanges}{{\color{red}#1}}{#1}}
\usepackage{tikz}
\usetikzlibrary{shapes.geometric, arrows, positioning, calc}

\DeclareUnicodeCharacter{2212}{-} 
\newcommand{\omite}[1]{}
\newcommand*{\missingreference}{\colorbox{red}{?reference?}}
\newcommand*{\missingcitation}{\colorbox{red}{?citation?}}
\makeatletter
\def\@setref#1#2#3{%
  \ifx#1\relax
   \protect\G@refundefinedtrue
   \nfss@text{\reset@font\missingreference}%
   \@latex@warning{Reference `#3' on page \thepage \space
             undefined}%
  \else
   \expandafter#2#1\null
  \fi}
\def\@citex[#1]#2{\leavevmode
  \let\@citea\@empty
  \@cite{\@for\@citeb:=#2\do
    {\@citea\def\@citea{,\penalty\@m\ }%
     \edef\@citeb{\expandafter\@firstofone\@citeb\@empty}%
     \if@filesw\immediate\write\@auxout{\string\citation{\@citeb}}\fi
     \@ifundefined{b@\@citeb}{\hbox{\reset@font\missingcitation}%
       \G@refundefinedtrue
       \@latex@warning
         {Citation `\@citeb' on page \thepage \space undefined}}%
       {\@cite@ofmt{\csname b@\@citeb\endcsname}}}}{#1}}
\makeatother 

\title{IoT-Zoo: A Container-Based Framework for Heterogeneous IoT Device Profiles and Reproducible Traffic Capture}

\author{
\IEEEauthorblockN{
    Vagner E. Quincozes\IEEEauthorrefmark{1},
    Diego Kreutz\IEEEauthorrefmark{2},
    Silvio E. Quincozes\IEEEauthorrefmark{2}
}
\IEEEauthorblockA{\IEEEauthorrefmark{1}Federal Fluminense University (UFF)}
\IEEEauthorblockA{\IEEEauthorrefmark{2}AI Horizon Labs and PPGES -- Federal University of Pampa (UNIPAMPA)}
}
    
\begin{document}

\maketitle

\begin{abstract} 
The validation of networking and security solutions for the Internet of Things (IoT) requires realistic and reproducible experimental data. However, existing platforms often achieve scalability by replicating a limited set of device types, which restricts profile diversity and fails to capture the heterogeneity of real-world IoT environments. In this paper, we present IoT-Zoo, a container-based testbed designed to support reproducible experimentation through heterogeneous, dataset-driven IoT device profiles. Built upon Containernet, IoT-Zoo automates the deployment of multi-domain scenarios and supports real application protocols such as MQTT and RTSP. The platform provides a single-command interface for environment provisioning and automated traffic capture (PCAP), enabling the generation of consistent traffic baselines and reducing the operational effort required to evaluate networking and security solutions.
\end{abstract}

\section{Introduction}

Internet of Things (IoT) research increasingly depends on reproducible experimental environments to evaluate networking, management, and security solutions under realistic traffic conditions~\cite{sah2025comprehensive,de2023survey}. However, building practical IoT testbeds remains challenging due to the inherent heterogeneity of real-world deployments, which vary significantly in terms of devices, application domains, and communication patterns~\cite{lima2025current, de2023survey,alex2023comprehensive}.

Most platforms address this challenge by scaling the number of emulated nodes, often reaching hundreds of devices. However, scale is frequently achieved by replicating only a few device types, which limits profile diversity and may underrepresent heterogeneous IoT behaviors (e.g., different sensing modalities, update rhythms, payload structures, and protocol stacks) \cite{mishra2025sdn,lima2025current,alex2023comprehensive}. In practice, scale is easier to achieve than heterogeneity.

This gap is clearly reflected in existing IoT testbeds and traffic generation frameworks, as summarized in Table~\ref{tab:iot_testbeds}. While several platforms achieve large-scale deployments, they do so with limited device diversity. For instance, Gotham reports 100 emulated devices but relies on only 11 distinct profiles, replicated to reach scale~\cite{gotham}. GothX extends this approach to 450 nodes, yet still inherits the same 11-profile structure~\cite{gothx}. Similarly, IoT-Flock models 18 devices using only four basic sensor types~\cite{iotflock}, and even large-scale generators such as STGen reach up to 6{,}000 nodes with only five sensor categories~\cite{islam2025stgen}. Across the literature, scale is consistently achieved by replication, whereas diversity remains constrained to a small set of profiles.

\begin{table}[!htb]
\scriptsize
\centering
\caption{IoT testbeds and traffic generation frameworks.}
\label{tab:iot_testbeds}
\begin{tabular}{@{} p{3cm} C{3.2cm} p{4.8cm} C{1.9cm} @{}}
\toprule
\textbf{Testbed / Project} & \textbf{Infrastructure} & \textbf{Diversity (Profiles \& Sensors)} & \textbf{Reproducibility} \\ 
\midrule

\textbf{IoT-Flock}\newline \cite{iotflock} & Virtual (Traffic Generator / GUI) & 4 sensor types (temperature, humidity, motion, light) & High \\ \midrule

\textbf{Automated IoT Testbed}\newline \cite{waraga2020design} & Physical & 2 device types (1 camera and 1 smart bulb). & Low \\ \midrule

\textbf{MQTT DoS Testbed}\newline \cite{syed2020denial} & Physical / Virtual mixed & 5 sensor types (motion, air quality, clock, temperature, gas). & Low \\ \midrule

\textbf{TON\_IoT}\newline \cite{moustafa2021new} & Hybrid (SDN, NFV, Fog/Edge) & 7 simulated sensors (via Node-RED) + 3 physical devices. & Medium \\ \midrule

\textbf{DoS/DDoS MQTT}\newline \cite{alatram2023ddos} & Physical & 13 sensor types (analog and digital sensors). & Low \\ \midrule

\textbf{ICSSIM}\newline \cite{dehlaghi2023icssim} & Emulated / Physical & 1 industrial process (bottle-filling factory with 2 PLCs). & High \\ \midrule

\textbf{Gotham}\newline \cite{gotham} & Emulated (GNS3 + Docker/VMs) & 11 distinct profiles (e.g., smart home, motor, hydraulics) & High \\ \midrule

\textbf{GothX}\newline \cite{gothx} & Emulated (GNS3 + Docker) & 11 distinct profiles based on Gotham & High \\ \midrule

\textbf{STGen}\newline \cite{islam2025stgen} & Virtual (Central core / CLI) & 5 sensor types (temperature, humidity, camera, GPS, switch) & High \\ \midrule

\textbf{Physical Smart Office}\newline \cite{farag2025development} & Physical & 3 sensor types (climate, sound, air) + 3 smart device types. & Low \\ \midrule

\textbf{IoMT Security}\newline \cite{zachos2025anomaly} & Physical (Medical focus) & 1 sensor type (producing temperature and humidity). & Low \\ \midrule

\textbf{IoT-Zoo} \newline \textbf{Our Proposal} & Virtual / Hybrid-ready (Containernet \& Docker) & \textbf{43 distinct profiles} (e.g., environmental, smart building, industrial, smart agriculture, human-centric, cameras) & High \\ 

\bottomrule
\end{tabular}

Reproducibility is classified as \textbf{High} (software-based, automated deployment via open-source scripts/templates), \textbf{Medium} (requires robust enterprise virtual infrastructure), or \textbf{Low} (requires purchasing and assembling physical hardware).

\end{table}

This imbalance reveals a fundamental limitation: existing testbeds prioritize instance scale over device-level heterogeneity, underrepresenting the variability of real-world IoT ecosystems, which encompass diverse sensing modalities, temporal patterns, payload structures, and protocol stacks. As a result, many experimental environments fail to capture the richness and variability required for realistic evaluation of networking and security solutions.

We address this limitation with IoT-Zoo, a container-based framework that treats profile diversity as a first-class design principle. As shown in Table~\ref{tab:iot_testbeds}, IoT-Zoo provides 43 distinct IoT profiles, representing the highest diversity among the compared testbeds. In contrast, the second most diverse platform supports 13 profiles~\cite{alatram2023ddos}, meaning that IoT-Zoo achieves more than 3.3$\times$ higher diversity, corresponding to an increase of approximately 231\%. This substantial improvement shifts the focus from predominantly replication-based scalability to diversity-driven modeling, without compromising scalability or reproducibility.

IoT-Zoo achieves this through three key design elements: (i) a profile-first architecture that models IoT devices as dataset-driven, containerized profiles; (ii) an extensible catalog of heterogeneous profiles grounded in real data characteristics; and (iii) a time-bounded execution workflow that automates deployment, orchestration, and traffic capture. This combination enables the construction of experimental environments that are both scalable and intrinsically heterogeneous, better reflecting the complexity of real-world IoT deployments.

\section{Architecture and Components}
\label{sec_architecture}

The IoT-Zoo architecture is designed to enable reproducible, extensible, and data-driven emulation of heterogeneous IoT environments, with explicit emphasis on device-level diversity. Unlike existing approaches that achieve scale by replicating a small set of device types, IoT-Zoo treats heterogeneity as a first-class architectural concern, enabling the systematic integration and execution of a large number of distinct device profiles within a unified experimental pipeline.

\begin{figure}[!htb]
    \centering
    \includegraphics[width=1\linewidth]{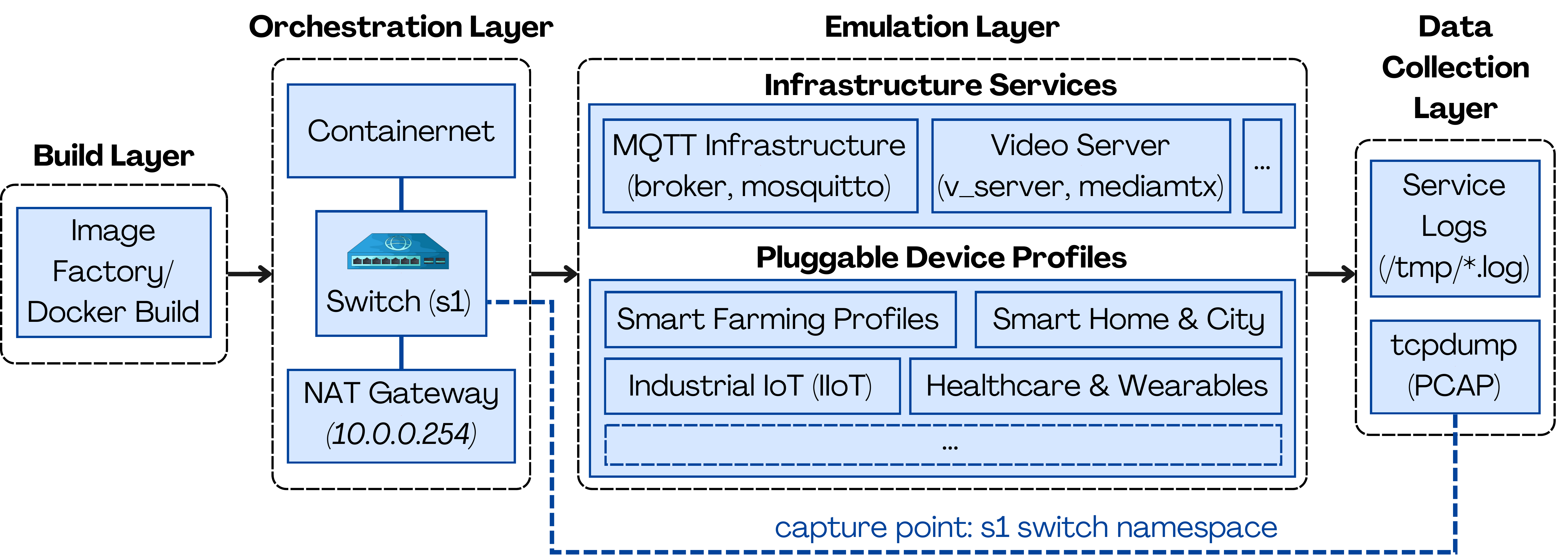}
    \caption{Architecture of the IoT-Zoo emulation testbed, illustrating the separation between build, orchestration, emulation, and data collection layers.}
    \label{fig:architecture}
\end{figure}

To support this goal, the architecture is organized into four layers: Build, Orchestration, Emulation, and Data Collection, as illustrated in Figure~\ref{fig:architecture}. These layers are loosely coupled and interact through well-defined interfaces, enabling modular evolution while preserving a consistent execution model.

In IoT-Zoo, heterogeneity is defined by the number and diversity of device profiles, rather than by the number of replicated instances. Each profile encapsulates a unique combination of data source, temporal dynamics, protocol stack, and payload structure. This abstraction allows the system to represent diverse IoT behaviors in a composable and extensible manner.

The \textit{Build Layer} is responsible for preparing all containerized components required by the testbed. It implements a Docker-based image factory where each device profile and infrastructure component is packaged as an independent image. This includes not only application logic but also dataset bindings and protocol configurations, enabling each profile to preserve its behavioral semantics across executions. By encapsulating profiles as self-contained artifacts, this layer supports consistent deployment and facilitates the dynamic inclusion of new device types without affecting existing components.

The \textit{Orchestration Layer} manages the lifecycle of experiments and the logical network topology. It is implemented using Containernet\footnote{\url{https://containernet.github.io/}}, which enables Docker containers to be deployed as network nodes interconnected through a virtual switch. Beyond standard responsibilities such as container instantiation, IP assignment, and routing configuration, this layer enables the dynamic composition of heterogeneous scenarios by instantiating arbitrary combinations of device profiles. It also controls execution parameters such as startup order, experiment duration, and termination, ensuring reproducible and time-bounded experiments. A central switch (\texttt{s1}) aggregates traffic, while a NAT gateway provides controlled external connectivity.

The \textit{Emulation Layer} is the core of the diversity-driven design and comprises Infrastructure Services and Pluggable Device Profiles. While infrastructure components (e.g., MQTT brokers and streaming servers) provide stable shared functionalities, device profiles are modular and dataset-driven, enabling the representation of diverse IoT behaviors across multiple domains.

Each profile is implemented as a container that transforms raw telemetry data into protocol-compliant traffic (e.g., MQTT and RTSP) via a streaming pipeline that preserves temporal dynamics. This design enables dynamic heterogeneity, allowing new device types to be incorporated by simply adding new datasets and profiles, without modifying the underlying infrastructure. Consequently, IoT-Zoo supports continuous expansion of its profile space, overcoming the limitations of static or template-based approaches.

The \textit{Data Collection Layer} captures all emulation outputs in a consistent, analysis-ready format. Network traffic is collected at the virtual switch level via traffic mirroring within the \texttt{s1} namespace, ensuring full visibility of inter-device communication without impacting execution. Device-level logs are also collected and temporally aligned with packet traces, enabling precise correlation between application events and network behavior. This integrated view supports downstream tasks such as intrusion detection, traffic classification, and behavioral analysis.

Overall, the architecture of IoT-Zoo enables a shift from replication-based scalability to diversity-driven modeling. By combining containerized profiles, dataset-driven behavior, and dynamic orchestration, the system supports large-scale yet inherently heterogeneous experimental environments, better reflecting the complexity of real-world IoT deployments.

\section{Reference Implementation of the IoT-Zoo Tool}
\label{sec_implementation}

\begin{figure}[!htb]
    \centering
    \includegraphics[width=1\linewidth]{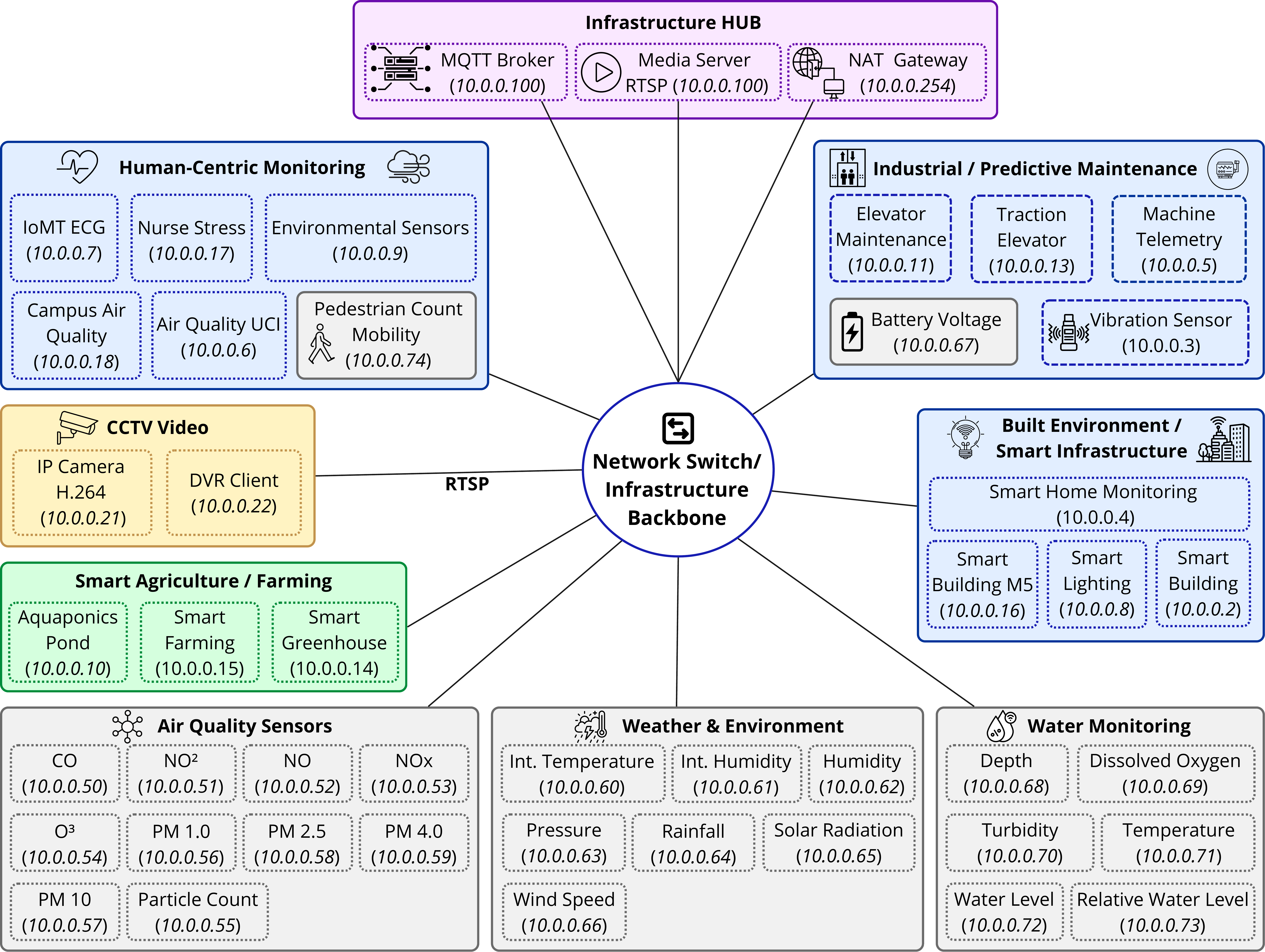}
    \caption{IoT-Zoo testbed topology highlighting device heterogeneity.}
    \label{fig:topology}
\end{figure}

Our reference implementation instantiates the proposed architecture using a star-tree topology centered on the virtual switch \texttt{s1}, as shown in Figure~\ref{fig:topology}. The deployment comprises 46 containers organized into functional categories and includes 43 distinct IoT device profiles, excluding infrastructure services such as the broker and media server. These profiles span multiple domains, including urban observability, human-centric monitoring, industrial and building automation, smart agriculture, and multimedia traffic.

The Urban Observatory-driven cluster (highlighted in grey in Figure~\ref{fig:topology}) forms the core of the emulation, using 2025 data from the Newcastle Urban Observatory\footnote{\url{https://urbanobservatory.ac.uk/}}. It integrates telemetry from 662 sensors, where containerized profiles act as aggregators (e.g., $NO_2$ from 34 stations and \textit{Pedestrian Count} from 218 sensors). This cluster generates continuous MQTT traffic representative of city-scale telemetry, including air quality, meteorology, water management, and mobility.

\begin{table}[ht]
\centering
\caption{Newcastle Urban Observatory subset (2025) for Smart City profiles.}
\label{tab:sensors}
\footnotesize

\begin{minipage}{0.49\linewidth}
\centering
\resizebox{\linewidth}{!}{
\begin{tabular}{llc}
\toprule
\textbf{Domain} & \textbf{Variable / Profile} & \textbf{Sources} \\
\midrule
\multirow{10}{*}{\textbf{Air Quality}} 
& Carbon Monoxide (CO) & 36 \\
& Nitrogen Dioxide (NO$_2$) & 34 \\
& Nitric Oxide (NO) & 34 \\
& Nitrogen Oxides (NO$_x$) & 28 \\
& Ozone (O$_3$) & 28 \\
& Particle Count & 26 \\
& Particulate Matter (PM 1.0) & 26 \\
& Particulate Matter (PM 10) & 20 \\
& Particulate Matter (PM 2.5) & 12 \\
& Particulate Matter (PM 4.0) & 8 \\
\midrule
\multirow{4}{*}{\textbf{Weather \& Indoor}} 
& Humidity & 38 \\
& Atmospheric Pressure & 31 \\
& Rainfall & 14 \\
& Wind Speed & 11 \\
\bottomrule
\end{tabular}
}
\end{minipage}
\hfill
\begin{minipage}{0.49\linewidth}
\centering
\resizebox{\linewidth}{!}{
\begin{tabular}{llc}
\toprule
\textbf{Domain} & \textbf{Variable / Profile} & \textbf{Sources} \\
\midrule
\multirow{3}{*}{\textbf{Weather \& Indoor}} 
& Internal Temperature & 9 \\
& Internal Humidity & 9 \\
& Solar Radiation & 2 \\
\midrule
\multirow{7}{*}{\textbf{Water}} 
& Water Level & 26 \\
& Relative Water Level & 25 \\
& Battery Voltage (Sensors) & 17 \\
& Water Temperature & 3 \\
& Turbidity & 3 \\
& Depth & 2 \\
& Dissolved Oxygen & 2 \\
\midrule
\textbf{Mobility} & Pedestrian Count & 218 \\
\midrule
\textbf{Total} & \textbf{Aggregated Physical Sensors} & \textbf{662} \\
\bottomrule
\end{tabular}
}
\end{minipage}

\end{table}

Beyond the Urban Observatory-driven core, IoT-Zoo incorporates additional profile categories to generate heterogeneous traffic patterns. Each containerized profile is driven by a distinct dataset, with representative examples discussed here due to the breadth of the instantiated profiles. For instance, Industrial / Predictive Maintenance includes profiles based on the AI4I 2020 Predictive Maintenance Dataset~\cite{matzka2020explainable}, while Human-Centric Monitoring includes physiological profiles such as IoTMT ECG from the mHealthDroid framework~\cite{banos2014mhealthdroid}. The remaining categories cover smart-building automation, smart agriculture / farming, and CCTV streams, combining lightweight MQTT telemetry with high-bandwidth RTSP traffic.

\section{Experimental Evaluation}

\subsection{Setup}

The evaluation was conducted in a virtualized environment (VMware Workstation Player) running Ubuntu 20.04.6 LTS with 4 vCPUs, 8 GB RAM, and 50 GB storage. The software stack includes Containernet (Mininet 2.3.1b1), Docker Engine 28.1.1, and Open vSwitch 2.13.8, enabling reproducible IoT emulation on commodity hardware.

\subsection{Methodology and Metrics}

Experiments were executed over a fixed duration of $T = 600$ seconds, with traffic captured at the virtual switch using \texttt{tcpdump}. Analysis was performed using \texttt{tshark}, focusing exclusively on MQTT \texttt{PUBLISH} packets to isolate application-level telemetry. TCP control traffic and keep-alive messages were excluded, and duplicate frames ($\Delta t < 100~\mu s$) were removed.

Three metrics were computed from packet arrival timestamps ($t_i$): Packet Count ($N$), Inter-Arrival Time ($IAT = \mu(\Delta t)$), and Jitter ($J = \sigma(\Delta t)$), capturing traffic volume, temporal behavior, and variability.

\subsection{Heterogeneity and Traffic Fidelity}

We quantify heterogeneity by analyzing per-profile message volume and temporal dynamics (IAT and jitter), as distinct IoT domains exhibit different periodicity and burstiness patterns. Table~\ref{tab:iot-metrics} summarizes the network metrics collected over a 600-second execution window. The results demonstrate the ability of IoT-Zoo to generate highly heterogeneous traffic profiles concurrently within the same topology, reflecting the diversity of real-world IoT environments. The analysis of inter-arrival times further reveals multiple coexisting behavioral patterns, ranging from stable periodic transmissions to bursty and irregular dynamics.

\begin{table}[ht]
\centering
\caption{Network metrics for representative IoT device profiles.}
\label{tab:iot-metrics}
\footnotesize
\begin{tabular}{lrrr | lrrr}
\toprule
\textbf{Device} & \textbf{Pkts} & \textbf{IAT(s)} & \textbf{Jit(s)} & \textbf{Device} & \textbf{Pkts} & \textbf{IAT(s)} & \textbf{Jit(s)} \\
\midrule
Building Rooms & 269 & 2.2111 & 2.5412 & Urban Air: PM10 & 2989 & 0.2003 & 0.3415 \\
Build: Int. Hum. & 301 & 1.9987 & 1.6460 & Urban Air: PM2.5 & 3055 & 0.1960 & 0.3390 \\
Build: Int. Temp. & 302 & 1.9857 & 1.6513 & Urban Air: PM4 & 2462 & 0.2433 & 0.4289 \\
Ind. Motor (Cooler) & 121 & 4.9588 & 0.4547 & Urban Air: Particle & 2450 & 0.2445 & 0.4297 \\
Legacy Air Sensor & 74 & 7.9919 & 6.2948 & Water L: Abs & 1812 & 0.3292 & 1.2396 \\
Mobility Sensor & 2292 & 0.2605 & 1.8041 & Water L: Rel & 1380 & 0.4322 & 1.4048 \\
Predictive Maint. & 120 & 4.9999 & 0.0993 & Water Q: DO & 180 & 3.3274 & 2.3576 \\
Smart Home (Domotic) & 78 & 7.6615 & 4.9212 & Water Q: Depth & 238 & 2.5134 & 2.5001 \\
Smart Lighting & 125 & 4.7996 & 0.9390 & Water Q: Temp & 208 & 2.8774 & 2.4701 \\
Urban Air: CO & 2338 & 0.2562 & 0.4364 & Water Q: Turb & 308 & 1.9403 & 2.4356 \\
Urban Air: NO & 762 & 0.7829 & 1.8172 & Weather: Hum & 302 & 1.9920 & 1.6473 \\
Urban Air: NO2 & 3519 & 0.1702 & 0.3228 & Weather: Press & 3843 & 0.1560 & 0.3623 \\
Urban Air: NOx & 597 & 0.9996 & 2.0002 & Weather: Rain & 1352 & 0.4411 & 1.4181 \\
Urban Air: O3 & 2894 & 0.2070 & 0.4050 & Weather: Solar & 592 & 1.0137 & 0.9991 \\
Urban Air: PM1 & 2999 & 0.1997 & 0.3411 & Weather: Wind & 1010 & 0.5940 & 0.8909 \\
e-Health ECG & 1184 & 0.5037 & 0.3834 & Human: Nurse Stress & 124 & 4.8572 & 1.0376 \\
Aquaponics & 117 & 5.1495 & 0.9955 & Smart Building M5 & 123 & 4.8955 & 0.9522 \\
Environmental Sensors & 121 & 4.9954 & 0.9436 & Farming Sensor & 120 & 4.9711 & 0.9409 \\
Greenhouse & 237 & 2.5146 & 1.3538 & Ind. Pred. Maint. GW & 119 & 5.0256 & 0.9107 \\
 & & & & Ind. Traction Elevator & 122 & 4.9423 & 0.9407 \\
\bottomrule
\end{tabular}
\end{table}

First, several industrial and infrastructure-oriented profiles, such as \textit{Ind. Motor (Cooler)}, \textit{Predictive Maint.}, \textit{Ind. Pred. Maint. GW}, and \textit{Ind. Traction Elevator}, exhibited near-periodic behavior, with IAT values close to $5.0s$ and low jitter. This confirms the testbed's ability to reproduce stable telemetry patterns typical of monitoring and predictive maintenance scenarios.

In contrast, Urban Observatory-driven profiles, represented by streams such as \textit{Weather: Press}, \textit{Urban Air: NO2}, and \textit{Urban Air: PM10}, generated sustained high-volume traffic. These devices produced between roughly 3000 and 3800 packets during the execution window, with low IAT values (down to $0.1560s$) and stable jitter, effectively stressing the network's throughput capacity.

Finally, profiles derived from more irregular real-world processes showed higher variability. For example, the \textit{Mobility Sensor} combined a low average IAT ($0.2605s$) with comparatively high jitter ($1.8041s$), reflecting bursty transmission dynamics associated with human activity. Other profiles, such as \textit{Building Rooms} and \textit{Water Q: Depth}, also exhibited higher timing dispersion, reinforcing that IoT-Zoo captures heterogeneous application behavior rather than relying on uniform synthetic traffic generation.

\subsection{Scalability \& Resource Efficiency}

To assess the feasibility of deploying heterogeneous IoT topologies on commodity hardware, we monitored the resource footprint of all 46 containerized profiles throughout the experiment. Table~\ref{tab:resources_domain} summarizes memory usage and CPU peak load grouped by the device categories defined in the instantiated topology. The results demonstrate that IoT-Zoo maintains a lightweight and predictable resource footprint even when combining diverse telemetry and multimedia workloads.

\begin{table}[ht]
    \centering
    \footnotesize
    \caption{Resource Consumption Profile by Device Category ($T=600s$).}
    \label{tab:resources_domain}
    \begin{tabular}{lccc}
        \hline
        \textbf{Device Category} & \textbf{Total Profiles} & \textbf{Mem. Usage (MiB) \textit{($\mu \pm \sigma$)}} & \textbf{CPU Peak (\%)} \\ \hline
        Air Quality Sensors & 10 & $48.65 \pm 1.98$ & $11.75$ \\ 
        Weather \& Environment & 5 & $48.65 \pm 2.02$ & $12.04$ \\ 
        Water Monitoring & 6 & $48.62 \pm 2.03$ & $11.26$ \\ 
        Built Environment / Smart Infra. & 8 & $20.82 \pm 16.24$ & $38.41$ \\ 
        Human-Centric Monitoring & 3 & $27.18 \pm 16.37$ & $7.71$ \\ 
        Industrial / Predictive Maintenance & 6 & $16.83 \pm 12.15$ & $7.67$ \\ 
        Smart Agriculture / Farming & 3 & $14.58 \pm 4.43$ & $9.93$ \\ 
        CCTV Video & 2 & $8.94 \pm 6.53$ & $5.27$ \\ 
        Infrastructure HUB & 3 & $12.44 \pm 10.01$ & $1.60$ \\ \hline
        \textbf{Total / Average} & \textbf{46} & \textbf{$\approx 27.42$ MiB} & - \\ \hline
    \end{tabular}
    \\[0.1cm]
\end{table}

The Urban Observatory-driven categories, namely \textit{Air Quality Sensors}, \textit{Weather \& Environment}, and \textit{Water Monitoring}, exhibited a highly consistent memory profile, with average usage close to $48.6$ MiB and low intra-category dispersion, with a standard deviation of approximately $2$ MiB. This uniformity across profiles processing different telemetry variables indicates that the stream-based ingestion approach keeps memory consumption bounded and predictable, regardless of the size of the underlying datasets.

The remaining categories showed lower average memory usage overall, including \textit{Industrial / Predictive Maintenance} ($16.83 \pm 12.15$ MiB), \textit{Smart Agriculture / Farming} ($14.58 \pm 4.43$ MiB), and \textit{CCTV Video} ($8.94 \pm 6.53$ MiB). \textit{Built Environment / Smart Infrastructure} displayed the highest variability and CPU peak ($38.41\%$), reflecting the coexistence of lightweight automation profiles and more heterogeneous building-oriented workloads within the same category.

Finally, the \textit{Infrastructure HUB} maintained a modest footprint ($12.44 \pm 10.01$ MiB) and low CPU peak ($1.60\%$), indicating that host resources are predominantly devoted to the emulated endpoints rather than orchestration overhead. Across the Urban Observatory-driven categories, CPU peaks remained around $11\%$--$12\%$, showing that data-driven replay can scale without saturating the allocated resources of a standard virtual machine (4 vCPUs).

\section{Availability and Demonstration Plan} \label{sec:availability}

To support validation and reuse, the IoT-Zoo source code, including Containernet scripts and Dockerfiles, is publicly available at \url{https://github.com/GT-IoTEdu/sbrc2026-IoT-Zoo}. The repository provides a \texttt{README} with Ansible-based installation and automated build instructions (\texttt{build\_images.sh}). The fully containerized testbed can be reproduced via a single configurable command (e.g., \texttt{sudo python3 run\_experiment.py --time 600}), enabling automated topology provisioning and traffic capture without manual intervention. An example of captured traffic in Wireshark is presented in Appendix~\ref{app:wireshark}.

For the SBRC Tools Lounge, the demonstration will run on a standard laptop (4 vCPUs, 8GB RAM) using an Ubuntu 20.04 VM, requiring no specialized hardware or network infrastructure. The demo will showcase automated setup, real-time testbed execution, and live traffic analysis with Wireshark, highlighting protocol and payload heterogeneity (JSON, XML, Binary).

\section{Conclusions and Future Work}\label{sec_conclusao}

This work introduced IoT-Zoo, a container-based testbed designed to enable reproducible and diversity-driven IoT experimentation. In contrast to existing platforms that rely on replication of a few device types, IoT-Zoo supports 43 distinct profiles within a unified pipeline, achieving the highest diversity among the evaluated testbeds. The architecture combines dataset-driven modeling, containerized execution, and automated orchestration to generate realistic multi-domain traffic with preserved temporal dynamics.

Experimental results show that heterogeneous patterns, from periodic telemetry to bursty human-driven behavior, can be reproduced concurrently with a lightweight and predictable resource footprint on commodity hardware. By bridging static datasets and live emulation, IoT-Zoo enables the systematic generation of reproducible datasets for traffic analysis, intrusion detection, and machine learning evaluation.

Future work will focus on increasing realism and applicability. We plan to incorporate benign user-driven behaviors, expand the profile catalog across domains, and integrate attack scenarios with ground-truth labeling. Additionally, we aim to couple IoT-Zoo with machine learning pipelines for end-to-end evaluation and to explore large-scale deployments with thousands of nodes.

\bibliographystyle{sbc}
{\bibliography{references}}

\appendix
\section{Example of Captured Traffic Output}
\label{app:wireshark}

\begin{figure}[!htb]
    \centering
    \includegraphics[width=1\linewidth]{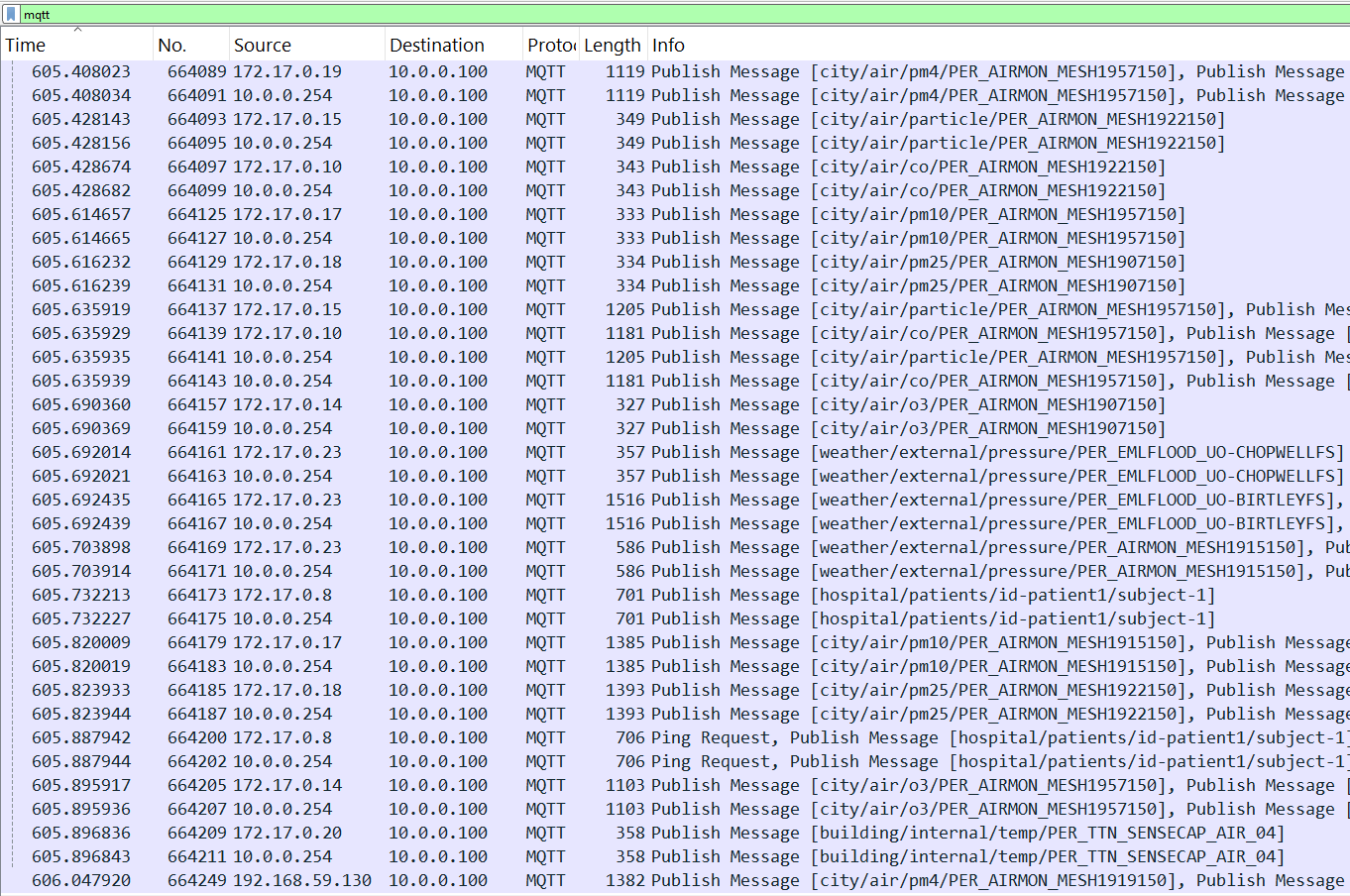}
    \caption{Example of MQTT traffic captured from an IoT-Zoo execution (PCAP inspection in Wireshark).}
    \label{fig:wireshark_example}
\end{figure}

Figure~\ref{fig:wireshark_example} shows an example of network traffic generated by IoT-Zoo and automatically captured during execution. The screenshot displays MQTT \texttt{PUBLISH} messages in Wireshark, highlighting heterogeneous topic namespaces and payload flows across multiple IoT domains, including Smart City telemetry (e.g., air quality), environmental monitoring (e.g., wind and pressure), e-Health (patient data streams), and industrial sensing (e.g., vibration). This output demonstrates the ability of IoT-Zoo to generate reproducible, multi-domain traffic traces in PCAP format, suitable for downstream analysis and benchmarking.

\begin{figure}[!htb]
    \centering
    \includegraphics[width=1\linewidth]{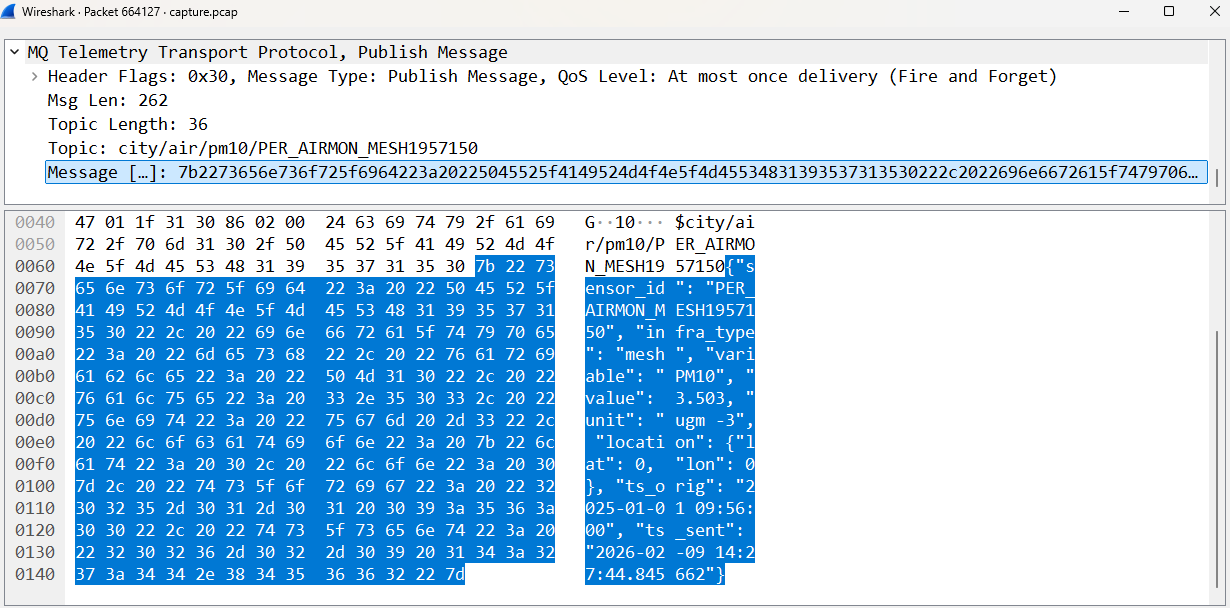}
    \caption{Decoded MQTT \texttt{PUBLISH} packet showing an example of the application payload replayed from the dataset.}
    \label{fig:wireshark_payload}
\end{figure}

Figure~\ref{fig:wireshark_payload} complements this view by presenting a decoded MQTT \texttt{PUBLISH} packet in Wireshark, illustrating an example of a structured application payload replayed from the original dataset. This demonstrates that IoT-Zoo preserves dataset-driven semantics at the application layer, rather than reproducing only protocol-level traffic patterns.

\end{document}